\documentstyle[aps,prl,psfig,special]{revtex}

\begin{document}
\draft
{
\title{Exact Study of the Effect of Level Statistics in Ultrasmall
Superconducting Grains
}
\author{G. Sierra${}^1$, J. Dukelsky${}^2$, G. G. Dussel${}^3$, 
Jan von Delft${}^4$, and Fabian Braun${}^4$
}
\address{${}^1$Instituto de Matematicas y Fisica Fundamental, C.S.I.C,
Madrid, Spain\\
${}^2$Instituto de Estructura de la Materia, C.S.I.C.,
Madrid, Spain\\
${}^3$Departamento de Fisica "Juan Jose Giambiagi", 
Universidad de Buenos Aires, Argentina\\
$^4$Institut f\"ur Theoretische Festk\"orperphysik, 
Universit\"at Karlsruhe, 76128 Karlsruhe, Germany
}
\date{August 31, 1999}
\maketitle
\begin{abstract}
  The reduced BCS model that is commonly used for ultrasmall superconducting
  grains has an exact solution worked out long ago by Richardson in the
  context of nuclear physics.  We use it to check the quality of previous
  treatments of this model, and to investigate the effect of level statistics
  on pairing correlations. We find that the ground state energies are on
  average somewhat lower for systems with non-uniform than uniform level
  spacings, but both have an equally smooth crossover from the bulk to the
  few-electron regime. In the latter, statistical fluctuations in ground state
  energies strongly depend on  the grain's electron number parity.
\end{abstract}
}
\narrowtext

Recent experiments by Ralph, Black and Tinkham, involving the
observation of a spectroscopic gap indicative of pairing correlations
in ultrasmall Al grains \cite{RBT}, have inspired a number of
theoretical \cite{vD,Braun1,BvD,SA,ML,MFF,BH,Braun2,DS,DP} studies of
how superconducting pairing correlations in such grains are affected
by reducing the grains' size, or equivalently by increasing its mean
level spacing $d \propto {\rm Vol}^{-1}$ until it exceeds the bulk gap
$\Delta$.  In the earliest of these, a grand-canonical (g.c.)  BCS
approach \cite{vD,Braun1,BvD} was applied to a reduced BCS Hamiltonian
for uniformly spaced, spin-degenerate levels; it suggested that
pairing correlations, as measured by the condensation energy $E^C$,
vanish abruptly once $d$ exceeds a critical level spacing $d^c$ that
depends on the parity (0 or 1) of the number of electrons on the
grain, being smaller for odd grains ($d^c_1 \simeq 0.89 \Delta$) than
even grains $(d^c_0 \simeq 3.6 \Delta$).  A series of more
sophisticated canonical approaches (summarized below) confirmed the
parity dependence of pairing correlations, but established
\cite{ML,MFF,BH,Braun2,DS,DP} that the abrupt vanishing of
pairing correlations at $d^c$ is an artifact of g.c.\ treatments:
pairing correlations do persist, in the form of so-called
fluctuations, to arbitrarily large level spacings, and the crossover
between the bulk superconducting (SC) regime $(d \ll \Delta)$ and the
fluctuation-dominated (FD) regime $(d \gg \Delta)$ is completely
smooth \cite{DS}. Nevertheless, these two regimes are qualitatively
very different \cite{Braun2,DS}: the condensation energy, e.g., is an
extensive function of volume in the former and almost intensive in the
latter, and pairing correlations are quite strongly localized around
the Fermi energy $\varepsilon_F$, or more spread out in energy,
respectively.

After the appearance of all these works, we became aware that the
reduced BCS Hamiltonian on which they are based actually has an exact
solution. It was published by Richardson in the context of nuclear
physics (where it is known as the ``picket-fence model''), in a series
of papers between 1963 and 1977 \cite{R1,R2} which seem to have
completely escaped the attention of the condensed matter community.
The beauty of this solution, besides its mathematical elegance, is
that it also works for the case of randomly-spaced levels.  It thus
presents us with two rare opportunities, which are the subject of this
Letter: (i) to compare the results of various previously-used
approximations against the benchmark set by the exact solution, in
order to gauge their reliability for related problems for which no
exact solutions exist; and very interestingly, (ii) to study the
interplay of randomness and interactions in a non-trivial model {\em
  exactly}, by examining the effect of level statistics on the SC/FD
crossover.

There is a previous study of the latter question by Smith and
Ambegaokar using the g.c.\ mean-field BCS approach \cite{SA}, who
concluded, interestingly, that randomness {\em enhances}\/ pairing
correlations: compared to the case of uniform spacings \cite{vD}, they
found that a random spacing of levels (distributed according to the
gaussian orthogonal ensemble) on average {\em lowers}\/ the
condensation energy $E^C$ to more negative values and increases the
critical level spacings at which $E^C$ vanishes abruptly, but these
still are parity dependent ($\langle d^c_1\rangle = 1.8 \Delta$,
$\langle d^c_0 \rangle \simeq 14 \Delta$).  However, the abrupt
vanishing of $E^C$ found by SA can be suspected to be an artifact of
their g.c.\ mean-field treatment, as was the case in
\cite{vD,Braun1,BvD}.  Indeed, our exact results for random levels
show (1) that the SC/FD crossover is as smooth as for the case of
uniformly-spaced levels; this means, remarkably, that (2) even in the
presence of randomness pairing correlations never vanish, no matter
how large $d/\Delta$ becomes; quite the opposite, (3) the
randomness-induced lowering of $E^C$ is strongest in the FD regime; in
the latter, moreover, (4) the statistical fluctuations in $E^C$ depend
quite strongly on parity.

{\em Exact solution.}--- Ultrasmall superconducting grains 
are commonly described \cite{vD,Braun1,BvD,SA,ML,MFF,BH,Braun2,DS,DP} 
 by a reduced BCS model, 
\begin{eqnarray}
  \label{eq:mod-hamiltonian}
  H = \sum_{j, \sigma= \pm} 
\varepsilon_{j\sigma} c_{j \sigma}^\dagger c_{j \sigma}
  -\lambda d \sum_{jj'}  c_{j +}^\dagger c_{j -}^\dagger 
c_{j' -} c_{j' +} \; ,
\end{eqnarray}
for a set of  pairs of time-reversed states $|j, \pm \rangle$
with energies $\varepsilon_j$, mean level spacing $d$ and
dimensionless coupling constant $\lambda$. 
Unbeknownst to the authors of \cite{vD,Braun1,BvD,SA,ML,MFF,BH,Braun2,DS,DP},
Richardson had long ago solved this model exactly, for 
\newpage 
\noindent an arbitrary set of levels $\varepsilon_j$ (not necessarily
all distinct): 
Since
{\em singly-occupied}\/ levels 
do not participate in and remain ``blocked'' \cite{Soloviev}
to the pairscattering
described by $H$,  
the labels of such levels 
are good quantum numbers.  
Let $|n,B\rangle$  thus denote 
an eigenstate with $N = 2n + b$ electrons, $b$ of which 
sit in a set $B$ of singly-occupied, blocked levels,
thereby contributing 
${\cal E}_B =
\sum_{i \in B} \varepsilon_i$ to the total energy.
The dynamics of the 
remaining $n$ pairs is then  governed by
\begin{equation}
 H_B = \sum_{j\not \in B} 2 \varepsilon_j b^\dagger_j b_j - \lambda \, d
\sum_{j,j' \not \in B} \; b_j^\dagger b_{j^{\prime}} \; ,
 \label{1}
\end{equation}
where  the pair operators $b_j = c_{j -} c_{j +} $  satisfy
``hard-core boson'' commutation relations, $[ b_j, b^\dagger_{j^{\prime}}] =
\delta_{j j^{\prime}} (1 - 2 b^\dagger_j b_j)$, and
the  sums are  over all {\em unblocked} levels.
Richardson showed that the lowest-lying of 
the eigenstates $|n,B \rangle$ has an (unnormalized)  product form,
$$
|n,B \rangle_G = \prod_{i \in B} c_{i \sigma}^\dagger
\prod_{\nu = 1}^{n} \biggl( \sum_{j \not \in B} 
{b_j^\dagger \over 2\varepsilon_j - e_\nu } \biggr) 
|{\rm Vac} \rangle \; , 
$$
%
where the $n$  parameters $e_\nu$ ($\nu = 1, \dots , n$) are that
particular solution of the $n$ coupled algebraic
equations 
\begin{equation}
\frac{1}{\lambda d} + \sum_{\mu=1 ( \neq \nu)}^{n} \frac{2}{ e_\mu - e_\nu} 
= \sum_{j \not \in B} \frac{1}{2 \varepsilon_j - e_\nu} \; , 
\label{4}
\end{equation}
%
that yields the lowest value for the ``pair energy''
${\cal E} (n) = \sum_{\nu =   1}^{n} e_\nu$.
Moreover, $|n, B \rangle_G$ has total energy  
$ {\cal E} (n)+ {\cal E}_B   $. The lowest-lying of all eigenstates 
with $n$ pairs and $b$ blocked levels, 
say $|n,b \rangle_G$ with energy ${\cal E}^G_b (n)$,
is that $|n, B \rangle_G$ for which the blocked
levels  in $B$ are all as close as possible to $\varepsilon_F$,
the Fermi energy of the  uncorrelated 
$N$-electron Fermi sea $|F_N\rangle$. 
In this Letter we shall always take all the $\varepsilon_j$
to be non-degenerate. The $e_\nu$, which may be thought of
as self-consistently-determined pair energies, then
coincide at $\lambda =0$ with the lowest $n$ energies
$2 \varepsilon_j$ ($j = 1, \dots, n$), and smoothly
evolve toward lower values as $\lambda$ is turned on.
This fact can be exploited during the 
 numerical solution of (\ref{4}),
which can be simplified by first making some
algebraic transformations, discussed in detail in \cite{R2}, 
 that render the equations less singular.

{\em Uniformly-spaced levels.}---
Our first application of the exact solution is to check
the quality of results previously obtained by various
other methods. Most previous works 
\cite{vD,Braun1,BvD,ML,MFF,BH,Braun2,DS} studied  a
half-filled band 
with fixed width $ 2\omega_D$ of uniformly-spaced levels (i.e.\ $\varepsilon_j
= j \, d$), containing $N=2n+b$ electrons.  Then the level spacing is $d= 2
\omega_D /N$ and in the limit $d \to 0$ the bulk gap is $\Delta = \omega_D
\sinh (1/\lambda)^{-1}$. Following \cite{Braun2}, we take $\lambda = 0.224$
throughout this paper.  To study the SC/FD crossover, two types of quantities
were typically calculated as functions of increasing 
$d/\Delta$, which mimics decreasing
grain size:  the even and odd ($b=0,1$) condensation energies
\begin{equation}
  \label{eq:condensation}
E^C_b (n) = {\cal E}^G_b (n) -
\langle F_N| H |F_N\rangle \; ;
\end{equation}
\begin{figure}
\centerline{\psfig{figure=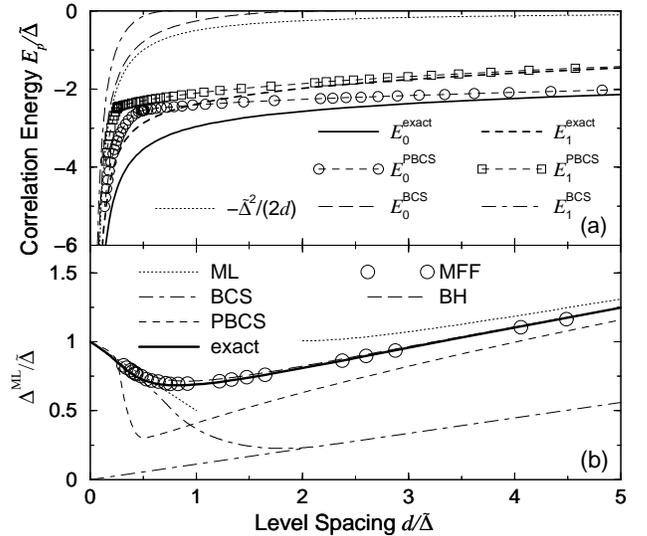,width=0.95\linewidth}} 
  \caption[Exact ground state condensation energies]{
    (a) The even and odd  $(b=0,1)$ condensation energies ${E}^C_b$ of
    Eq.~(\ref{eq:condensation}),
   calculated with BCS, PBCS and exact wave
    functions, as functions of $d/  \Delta = 2 \sinh (1/ \lambda) / (2n
    + b)$, for $\lambda = 0.224$.  For comparison the dotted line gives the
    ``bulk'' result $E_0^{\rm bulk} = -\Delta^2/(2d)$.  (b) Comparison
    of the parity parameters $\Delta^{\text{ML}}  $ \protect\cite{ML} 
   of Eq.~(\ref{eq:ML}) obtained
    by various authors: ML's analytical result (dotted lines)
    [$\Delta(1-d/2 \Delta)$ for $d\ll \Delta$, and $d/2\log(a
    d/\Delta)$ for $d\gg \Delta$, with $a=1.35$ adjusted to give
    asymptotic agreement with the exact result]; grand-canonical
    BCS approach (dash-dotted line) [the naive perturbative result
    $\frac12\lambda d$ is continued to the origin]; PBCS approach
    (short-dashed line); Richardson's exact solution (thick solid line); exact
    diagonalization and scaling by MFF (open circles) and BH (long-dashed
    line).  }
  \label{fig:exact-gse}
\end{figure}
\noindent
and a parity parameter introduced by Matveev and Larkin (ML) \cite{ML} to
characterize the even-odd ground state energy difference,
\begin{equation}
  \label{eq:ML}
\Delta^{\text{ML}} (n)= {\cal E}^G_1(n) - 
[ {\cal E}^G_0(n) + {\cal E}^G_0 (n+1)]/2 \, .
\end{equation}
Following the initial g.c.\ studies \cite{vD,Braun1,BvD,SA,ML},
the first canonical study  was that of Mastellone, Falci and Fazio (MFF)
\cite{MFF}, who used Lanczos exact diagonalization (with $n \le 12$)
 and a scaling argument to probe the crossover regime.  Berger and
 Halperin (BH) \cite{BH}
showed that essentially the same results could be
 achieved with $n \le 6$ by first reducing the bandwidth and renormalizing
 $\lambda$, thus significantly reducing the calculational effort
involved.  To access larger systems and fully recover the bulk limit,
 fixed-$n$ projected variational BCS wavefunctions (PBCS) were used in
\cite{Braun2} (for $n \le 600$); significant improvements over the
 latter results, in particular in the crossover regime,
 were subsequently achieved in \cite{DS}
 using the density matrix renormalization group (DMRG) (with $n \le 400$).
 Finally, Dukelsky and Schuck \cite{DP} showed that 
a self-consistent RPA approach, that in principle can
be extended to finite temperatures, describes the f.d.\ regime
rather well (though not as well as the DMRG).

To check the quality of the above methods,
we \cite{Braun-thesis} computed  $E_b^C(n)$ and 
$\Delta^{\rm ML}(n)$ using Richardson's solution 
(Fig.~\ref{fig:exact-gse}). 
 The exact results
(a)
quantitatively agree, for $d \to 0$, with 
the leading $- \Delta^2/2d$
behavior for $E^C_b(n)$ obtained in 
the g.c.\ BCS approach \cite{vD,Braun1,BvD}, which in 
this sense is exact in the bulk limit, corrections being of order $d^0$;
(b) confirm
 that a completely smooth \cite{DS} 
crossover occurs around the scale $d \simeq
 \Delta$ at which the g.c.\ BCS approach breaks down;
(c) show that the PBCS crossover \cite{Braun2}
 is qualitatively correct, but not 
quantitatively, being somewhat too abrupt;
(d) are reproduced remarkably well by the approaches of MFF \cite{MFF} and BH
\cite{BH}; (e) are fully reproduced by the DMRG of \cite{DS} with a relative
error of $< 10^{-4}$ for $n \le 400$; our figures don't show DMRG curves,
since they are indistinghuishable from the exact ones and are discussed in
detail in \cite{DS}.

The main conclusion we can draw from these comparisons is
that the two approaches based on renormalization group ideas
work very well: the DMRG is essentially exact for this model,
but the band-width rescaling method of BH also gives
remarkably (though not quite as) good results with rather less effort.
In contrast, the PBCS approach is rather unreliable in the
crossover region. 

\begin{figure}
\centerline{\psfig{figure=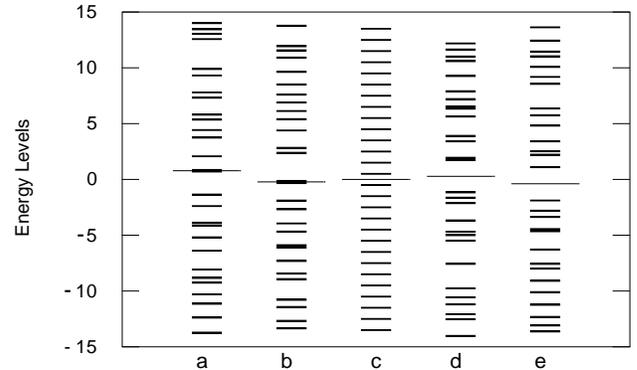,width=0.95\linewidth}} 
\caption{Sets of energy levels with $N = 28$. 
Set $c$ has equally spaced levels, with spectroscopic gap
[Eq.~(\ref{spectroscopic})] $E^S_0/d = 1.54$.  
Sets $a,b$ (or $c,d$) are randomly spaced;
among all sets with $N=28$ we studied, 
the ones shown have the
smallest (largest) values for  $E_{0}^{S}/d$,
namely  0.886, 0.891 $(3.30, 3.37)$, 
due to the small (large) spacing between the two
levels closest to  $\varepsilon_F$,  
illustrating how random level fluctuations affect energy gaps. 
}
\label{fig1}
\end{figure}

{\em Randomly-spaced levels.}--- The remainder of this Letter
addresses the question of how randomness of the levels $\varepsilon_j$
affects pairing correlations.  We studied half-filled bands of $N = 2n
+ b$ non-uniformly spaced but non-degenerate levels (for $N \le 260$),
with $b=0,1$.  The energy levels in small metallic grains with time
reversal symmetry follow the gaussian orthogonal ensemble distribution
\cite{GSO}. We generated sets of levels $\varepsilon_i$ $(i=1,
\cdots,N)$ by diagonalizing $ 2N \times 2N$ random matrices, taking
$N$ adjacent values from the central part of the eigenspectrum
(to avoid boundary effects) and performing the rescaling \cite{SA}
\begin{equation}
\varepsilon \rightarrow \frac{1}{2 \pi} \left[ 4N \; {\rm sin}^{-1}
\left( \varepsilon / \sqrt{4 N} \right) + \varepsilon \sqrt{ 4 N -
\varepsilon^2} \right] \; , 
  \label{15}
\end{equation}
to ensure an average level spacing of
one in units of $d$. In Figure~\ref{fig1} we show four such sets of
randomly generated levels for $N = 28$, together with the equally
spaced set.

\begin{figure}
\centerline{\psfig{figure=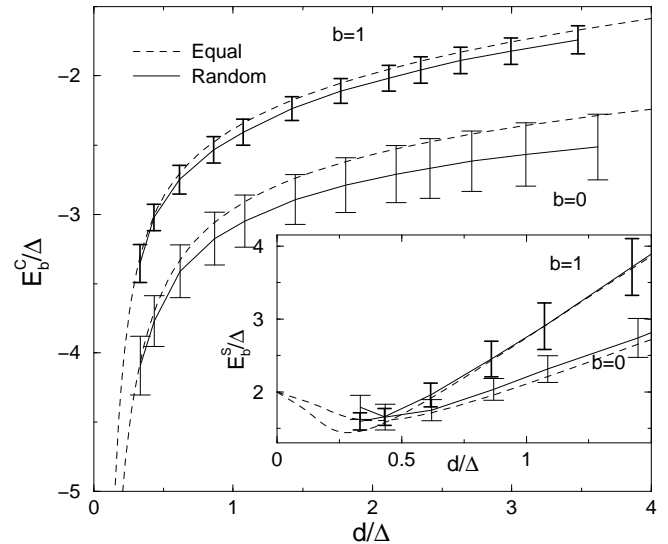,width=0.98\linewidth}} 
\caption{Exact even and odd condensation energies,
$E^C_b$  
for equally spaced
levels  (dashed line), and the ensemble-average 
$\langle E^C_b \rangle$ for randomly-spaced levels (solid line). 
The height of the fluctuation bars gives the variances $\delta E^C_b$.
The inset shows the corresponding spectroscopic gaps $E^S_b$
and variances $\delta E^S_b$.}
\label{fig2}
\end{figure}

For each such set of $2n+b$ levels,  we calculated 
the exact ground state energy 
${\cal E}^G_b (n)$, the condensation energy $E^C_b(n)$, and the spectroscopic
gap  \cite{BvD}
\begin{equation}
\label{spectroscopic}
E_b^S(n) = {\cal E}^G_{b+2} (n-1) - {\cal E}^G_b(n), $$
\end{equation}
which gives the energies needed to break a single pair in the (even or
odd) ground state.  Subsequently we calculated the ensemble
average $\langle E^C_b (n)\rangle $ and variance $\delta E^C_b (n) =
\sqrt {\langle(E^C_b)^2\rangle - \langle E^C_b\rangle ^2 }$ (and
analogously $\langle E_b^S \rangle $ and $\delta E_b^S$) over many
realizations of random matrices.  The ensemble size was 1000 for $24
\leq N \leq 40$, and varied between 700 and 150 for $40 \leq N \leq
260$.  Figure~\ref{fig2} presents our results for these
ensemble-averages (solid lines, with variances indicated by
fluctuation bars) together with those for the uniformly-spaced (u.s.)
set discussed above (dashed lines).  It shows a number of interesting
features. 

Firstly, the
two main conclusions of SA \cite{SA} are confirmed, namely (a) that
pairing correlations are on average {\em stronger}\/ for randomly-
than for uniformly-spaced levels, $\langle E_b^C \rangle < E_b^C {\rm
  (u.s.)}$; and (b) that the parity effect persits in the presence of
randomness, $\langle E_0^C \rangle < \langle E_1^C \rangle $.  In SA's
g.c.\ calculation these facts could be understood \cite{SA} from a
condition, derived from the BCS gap equation,
for having non-vanishing pairing correlations, namely
$2/\lambda < \sum_{j \not \in B} 1 / |\bar \varepsilon_j - \bar \mu|.$
Here $\bar \varepsilon_j$ and
the g.c.\ chemical potential $\bar \mu$ are in units of $d$,
and the number of terms in the sum is of order $2 \omega_D/d$.
As $d$ increases, this number decreases, until
the inequality ceases to hold at a critical spacing $d^c_b$.
Since statistical fluctuations to smaller values of
$|\bar \varepsilon_j - \bar \mu|$
 carry more weight than those to larger values, fluctuations on
average tend to increase $d^c_b$, which explains (a);
moreover, since
the blocking of levels close to $\bar \mu$ reduces 
the number of terms in the sum, it reduces $d_b^c$,
which explains (b). 

Since the equation on which SA's elegant argument is based 
breaks down in the FD regime, let us attempt
another way of interpreting (a) and (b): pairing correlations
involve a non-zero amplitude to find pair states with $\varepsilon_j >
\varepsilon_F$ doubly occupied and ones with $\varepsilon_j < \varepsilon_F$
empty.  Such correlations between states below and above $\varepsilon_F$,
called ``pair-mixing across $\varepsilon_F$'' in \cite{vD}, gain interaction
energy but cost some kinetic energy.  The latter cost is the smaller, the
closer the states involved in pair-mixing across $\varepsilon_F$ lie together
(which is why the bulk limit $d \to 0$ is so strongly correlated).
Statistical fluctuations in level positions that yield more-closely or
less-closely spaced levels around $\varepsilon_F$ than for the uniform case,
would thus cause a respectively lower or higher kinetic energy cost for
pairmixing across $\varepsilon_F$; according to (a), the former on average
outweighs the latter, just as had SA concluded in \cite{SA}.
Furthermore, in odd grains the blocked level at $\varepsilon_F$ always
causes the spacing between pair levels below and above
$\varepsilon_F$, and hence the kinetic energy
cost for pair-mixing across $\varepsilon_F$, to be somewhat
larger than in even grains, which explains (b).

Now, the ability of the exact solution to correctly treat the FD
regime enables us to uncover several further facts that are beyond the
reach of SA's g.c.\ mean field approach: (c) The SC/FD crossover is as
smooth for randomly- as for uniformly-spaced levels, confirming that
the abrupt vanishing of pairing correlations at some critical level
spacing found by SA is an artifact of their g.c.\ mean field
treatment, just as in \cite{vD,Braun1,BvD}.  (d) Even in the presence
of randomness, pairing correlations never vanish, no matter how large
$d / \Delta$.  Quite the opposite, (e) the randomness-induced lowering
in condensation energy to more negative values, $\langle E_b^C \rangle
- E_b^C {\rm (u.s.)}$, is {\em strongest}\/ in the FD regime; this
perhaps somewhat counterintuitive result illustrates that the smaller
the number of levels is that lie ``close to'' (i.e.\ within $\Delta$
of) $\varepsilon_F$, the stronger is the effect of fluctuations in
their positions on the kinetic energy cost for pair-mixing;
conversely, this randomness-induced lowering of $E^C_b$ decreases in
the crossover regime and becomes negligible in the SC regime, in which
very many levels lie within $\Delta$ of $\varepsilon_F$.  (f) The
variances $\delta E^C_b$ are essentially $d$-independent in the range
$24 \leq N \leq 260$, implying that the {\em
  relative}\/ statistical fluctuations of $E^C_b$ should be negligible
in the bulk limit, as expected.

Remarkably, we can also discern (g) three ``parity-dependent fluctuation
effects'', in that the following three quantities are larger for even than for
odd grains: (g1) the variances $\delta E^C_b$ (with $\delta E^C_0 \simeq 2 \,
\delta E^C_1 \simeq \Delta/2$); and the randomness-induced changes in (g2)
condensation energies $|\langle E_b^C \rangle - E_b^C {\rm (u.c)}|$ and (g3)
spectroscopic gaps $|\langle E_{b}^{G}\rangle -E_{b}^{G}({\rm u.s.})|$ (inset
of Fig.~\ref{fig2}). All three of these effects have the same origin as the
more familiar parity effect (b), namely blocking: the more levels around
$\varepsilon_F$ are blocked, the larger the effective spacing between 
states involved in pair-mixing across $\varepsilon_F$, 
and hence the smaller the sensitivity of the
total energy to statistical fluctuations in level positions.

In conclusion, using  Richardson's exact solution
we have found that level randomness does
not modify the smooth nature of the SC/FD crossover.
It just enhances pairing correlations  somewhat
compared to those of uniformly-spaced levels,
having the  strongest effect in the FD regime.
In the latter we found 
that statistical fluctuations become strongly parity dependent.

{\em Acknowledgements.}--- We thank R.\ Richardson for al\-lerting us
to his work, and V.\ Ambegaokar, F.\ Evers, and P.\ Schuck
for discussions.  This work was supported by the DGES grants
PB95-01123 (J.D.) and PB97-1190 (G.S.), and by ``SFB 195'' of the DFG
(J.v.D. and F.B.). \vspace*{-8mm}



\widetext
\end{document}